\documentclass[%
aip,amsmath,amssymb, reprint
]{revtex4-1}

\usepackage{graphicx}
\usepackage{dcolumn}
\usepackage{bm}
\usepackage[ruled,vlined]{algorithm2e}
\usepackage[utf8]{inputenc}
\usepackage[T1]{fontenc}
\usepackage{mathptmx}
\usepackage{listings}
\usepackage{tikz}
\usepackage{natbib}
\setcitestyle{square, comma, numbers,sort&compress, super}
\usepackage{cancel}
\usetikzlibrary{arrows}
\usetikzlibrary{trees}
\usepackage{midfloat}
\usepackage{cancel}
\usetikzlibrary{decorations.pathmorphing}
\usetikzlibrary{decorations.markings}
\usetikzlibrary{automata,positioning}
\usepackage{float}
\usepackage{caption}
\usepackage{simplewick}
\usepackage{csquotes}
\usepackage{graphicx}
\usepackage{subcaption}
\usepackage{multirow}
\usepackage{cleveref}
\usepackage{amsmath}
\usepackage{tabularx}
\usepackage{natbib}
\usepackage{xcolor,color,soul}
\usepackage{xcolor}
\definecolor{codegreen}{rgb}{0,0.6,0}
\definecolor{codegray}{rgb}{0.5,0.5,0.5}
\definecolor{codepurple}{rgb}{0.58,0,0.82}
\definecolor{backcolour}{rgb}{0.95,0.95,0.92}
\lstdefinestyle{mystyle}{
    backgroundcolor=\color{backcolour},   
    commentstyle=\color{codegreen},
    keywordstyle=\color{magenta},
    numberstyle=\tiny\color{codegray},
    stringstyle=\color{codepurple},
    basicstyle=\ttfamily\footnotesize,
    breakatwhitespace=false,         
    breaklines=true,                 
    captionpos=b,                    
    keepspaces=true,                 
    numbers=left,                    
    numbersep=5pt,                  
    showspaces=false,                
    showstringspaces=false,
    showtabs=false,                  
    tabsize=2
}

\sethlcolor{yellow}
\begin{document}

\preprint{AIP/123-QED}

\author{Anish Chakraborty} \affiliation{Department of Chemistry, Indian Institute of 
Technology Bombay, Powai, Mumbai 400076, India}  
\author{Rahul Maitra}\email{rmaitra@chem.iitb.ac.in} 
 \affiliation{Department of Chemistry, Indian Institute of Technology Bombay, Powai, 
 Mumbai 400076, India}
\title{
Fixing the Catastrophic Break-down of Single Reference Coupled Cluster Theory for 
Strongly Correlated Systems: Two Paradigms towards the Implicit Inclusion of High Rank
Correlation with Low-Spin Channels}

\begin{abstract}

The dual exponential coupled cluster (CC) theory proposed
by Tribedi \textit{et al.}[J. Chem. Theory Comput. 2020, 16, 10, 
6317–6328] performs significantly better for a wide range 
of weakly correlated systems than the
coupled cluster theory with singles and doubles excitations
(CCSD) due to the implicit inclusion of high-rank excitations. The
high-rank excitations are included through the action of a 
set of vacuum annihilating scattering operators that act 
non-trivially on certain correlated wavefunctions and are
determined \textit{via} a set of local denominators involving the energy
difference between certain excited states. This often leads the 
theory to be prone to instabilities. In this manuscript, we show
that restricting the correlated wavefunction, on which the scattering
operators act upon, to be spanned by only the singlet paired
determinants can avoid the catastrophic breakdown.
For the first time, we present two nonequivalent approaches to 
arrive at the working equations, \textit{viz.} the projective 
approach with sufficiency conditions and the amplitude form with 
many-body expansion. While the effect of the triple 
excitation is quite small around molecular equilibrium geometry,
this scheme leads to a better qualitative description
of the energetics in the regions of strong correlation.
With a number of pilot numerical applications, 
we have demonstrated the performance of the 
dual-exponential scheme with both the proposed 
solution strategies while restricting the 
excitation subspaces coupled to the corresponding
lowest spin channels.

\end{abstract}

\maketitle

\section{Introduction}
Single reference Coupled Cluster (CC) theory\cite{cc1,cc2,cc3,cc4} 
has been the method of choice for a balanced description of accuracy and
affordability for small molecular systems.
While one would ideally resort to the Full Configuration 
Interaction (FCI) method for an 
exact solution, it is extremely expensive with a $N!$ scaling, 
where $N$ is the number of basis functions. For weakly to
moderately correlated systems, CC theory with singles and 
Doubles excitation (CCSD) scheme performs quite well 
compared to the FCI solutions. The CCSD scheme with 
perturbative triples correction, 
CCSD(T)\cite{ccsdpt1,ccsdpt2,ccsdpt3} provides a balanced
description of computational cost and accuracy, and is 
termed as the Gold-Standard in quantum chemistry. Although 
the CC scheme with SD and iterative triples (CCSDT) scheme 
is in principle more exact, however, it comes with a higher
computational scaling which often make the theory inaccessible
to treat larger molecular systems. The advantage with CC theory 
is due to the size-extensive and size-consistent description of 
molecular energetics upon any order of truncation of the 
rank of the cluster operators.

Recently, a CC theory (termed as the iterative n-body excitation
inclusive CCSD (\textit{iCCSDn})) with iterative inclusion of triple
excitation has been developed by the present authors 
\cite{iccsdn1,maitra_coupled_2017,maitra_correlation_2017}
that incorporates high-rank excitations in a non-perturbative
manner, with a computational scaling of the same order as that
of conventional CCSD. The authors designed a novel double 
exponential ansatz comprising
of two sets of operators, namely, the cluster operators, $T$,
inducing hole to particle excitations, and a vacuum annihilating
scattering operator, $S$, that acts
on correlated ground state.
The structure of the scattering operators are particularly
interesting (\textit{vide infra}) as it contains quasi-hole/
quasi-particle destruction component in it, and thus it can
selectively act upon only certain excited determinants
that span the correlated ground state wavefunction, while it annihilates
the Hartree-Fock (HF) vacuum. The 
component of $S$ that has quasi-hole destruction operator is denoted 
as $S_h$ and the same with quasi-particle destruction operator is termed as 
$S_p$. Moreover, since the action of $S$ (either 
$S_h$ or $S_p$) annihilates the HF vacuum, one cannot derive a closed 
form of equations to determine them; rather they are determined via a 
many-body expansion with \enquote*{\textit{local energy denominator}} 
that involves the energy difference between the excited determinant 
it acts upon and the excited state determinant it leads to.
Due to the energy difference of two excited determinants in the denominator of the
amplitude determining equations, the description of $S$ often
leads to instability. In this manuscript, we investigate how one
may avoid this failure of iCCSDn by restricting the doubly 
excited subspace on which the $S$ acts to their low spin 
coupled counterparts. One may note that at this point, 
that the local treatment of terms which leads to
increasingly high rank excitations is conceptually
different from the many-body perturbation theory
(MBPT). In case of MBPT, at each order, the global
denominator involves the energy difference between 
the excited determinant and the ground HF determinant.
While an amplitude formulation by explicit many-body 
expansion is now well understood\cite{nooijen2000,iccsdn1}, 
there is as such not much literature on how
a rank-two generalized \enquote{excitation operator} can be 
determined by projecting against excited determinants. 
In the current framework, we develop two inequivalent
paradigms towards the genesis of the working equations:
\begin{enumerate}
\item For the first time, to the best of our knowledge,
we develop a projective formulation towards the determination 
of the amplitudes corresponding to the 
scattering operators. Note that due to the vacuum
annihilating property of the scattering operators, 
or any hole-particle generalized \enquote{excitation 
operators} in that sense, a closed form of expression 
is not realizable. Thus, towards the determination of 
the $s-$amplitudes, we adopt the strategy of 
determining the corresponding matrix elements with respect to the
various rank-two and rank-three excited determinants. 
The redundancy that might appear in such cases are 
also avoided by invoking suitable sufficiency 
conditions as we will discuss in subsection \ref{projective}.
\item We also recapitulate the amplitude formulation 
where the double similarity transformed hamiltonian 
is expanded in various many-body terms, and the $s-$ 
and the $t-$ amplitudes are updated at each step with 
the many-body effective hamiltonian that is
structurally identical to the associated $S$ and $T$ 
operators using their local denominators. 
\end{enumerate}
As mentioned previously, in often cases of strongly correlated 
systems like bond dissociation, the cluster
amplitudes become ill-defined due to the 
underlying instability of the correlated state 
generated by the truncated CC expansion.
\cite{high_ex1,high_ex2,high_ex3,high_ex4,nCC,pCCSD,CCSD_VBT,distinguish_cluster1,distinguish_cluster2,distinguish_cluster3,distinguish_cluster4,poly_BCS}
This is thought to be originated from the vanishing energy gap 
between the occupied and virtual orbitals and thus any low order perturbation 
theory is plagued due to the vanishing energy denominator. However, as Scuseria
\cite{singlet2} had pointed out, any non-perturbative theory, 
like CC, is likely to overcome this issue with
sufficiently high-rank parametrization of the wavefunction. However, this may not
be the sole reason: in the strong correlation limit, the restricted HF wavefunction 
(RHF) tends to be unstable to symmetry broken unrestricted HF solution. This implies 
that one needs to adapt to a set of good quantum numbers associated with the 
unrestricted HF (UHF) solution. This motivated Bulik \textit{et. al.}\cite{singlet2} 
to develop the singlet coupled CC theory where the spin symmetry of the excitation 
channels are preserved throughout.

The issue with small denominator may get even more pronounced for
the parent iCCSDn as it simulates the high rank excitations through a set of
local denominators. One may further note that the $S-$operators selectively act on 
certain correlated functions generated by the 
truncated CCSD expansion. Thus, if the CCSD description fails due to the
poor choice of the reference function, the iCCSDn would be unable to
bypass the catastrophic breakdown. Neither the projective nor the amplitude
formulation would likely be suitable in such cases.
One may wonder whether fixing the spin-symmetry of the function on which 
the $S$ operators act would make the theory more stable under strong correlation. This
leads us to the development of the current methodology where the functions on which the 
scattering operators act are adapted to a given spin symmetry. We would also 
demonstrate that such a restriction on the ground state 
wavefunction fixes the catastrophic breakdown of our theory while 
it encapsulates substantial high-rank dynamic correlation
that a spin-channel restricted CCSD counterpart fails to account for.
One may however note that in the current manuscript, although we have
imposed a restriction over the spin symmetry of the cluster operators, the
theoretical methods discussing two paradigms are entirely general. 
Furthermore, we comment that
this paper is by no means to be considered as a solution to the intruder state problem
for the reasons stated above.

The idea of imposing the restrictions on the possible 
spin-channels is not entirely new as there have been 
several approaches that have explored this idea. 
Historically the restriction of spin channel stems from pair
CCD (pCCD)\cite{pccd1,pccd2,pccd3,pccd4,pccd5,pccd6}
where only those rank-two excitations are 
considered where both the electrons are excited from the
same spatial orbitals. These excitations are screened 
in terms of their seniority numbers. While such restriction
on the double excitation operators brings in the 
stability of the CC solutions, as pointed out by Bulik \textit{et. al.}\cite{singlet2}, 
the pCCD formulation necessitates full orbital optimization.
The authors in Ref. [30], on the other hand, proposed an
alternate scheme with the good features of pCCD 
where part of the cluster operators based on their spin-coupling
channels was removed while retaining their eigenvector property
\cite{singlet1,singlet2,singlet3}. This is 
conceptually different from Random Phase Approximation (RPA)\cite{rpa1,rpa2,rpa3}
as pointed out by the authors where only a selection of the 
terms are retained in the CC effective hamiltonian whereas
the cluster operators are taken in full. With the condition that the
double excitation operators are restricted to their singlet 
coupled channel, the resulting CCD0 fixes the catastrophic 
failure of conventional CCD. One may note that a similar analysis 
was performed by one of the present authors in the context of 
second order perturbation theory where the author dynamically 
tuned the contributions of the singlet and triplet paired channels 
of strongly correlated molecules over their potential energy surface
and demonstrated that such a scheme can qualitatively explain the 
bond dissociation even within perturbation theory framework.\cite{rm_scsmp2}
In this article, we had taken the route developed by Scuseria 
by selectively including only those cluster operators which 
are coupled to a singlet channel. 
However, one still needs to replenish
the correlation that is missed out at the two body level. 
The correlation lost due to exclusion of the triplet spin channels 
was compensated by inclusion of the triple excitations. 
The triple excitations were
included in an iterative manner via the projective and amplitude
equations where the two-body cluster amplitudes are restricted to
its singlet coupled counterparts while the scattering operator 
is spin-unrestricted. Due to the spin restriction on the two-body 
cluster operators, the higher-body excited terms generated by the action of the 
scattering operators on the singlet-coupled states are also low-spin coupled
functions.

In this manuscript, we begin with the general structure of the 
dual-exponential ansatz in Sec. \ref{dualexp}, followed by the 
discussion on the choice of the singlet paired rank-two cluster 
operators in Sec. \ref{singletT}. With the choice of the operators
that enter in the wavefunction parametrization, we discuss in 
subsec. \ref{projective} 
the solution paradigm-I on the explicit projective structure of the 
working equation and discuss how one can impose a physically appealing
sufficiency condition to bypass the redundancy to determine the 
scattering amplitudes. In subsec. \ref{amplitude}, we recapitulate the 
amplitude formulation of our theory where in the context of the 
manuscript, the effective hamiltonian residues are formed for the 
singlet paired channels. Following the development of the two 
non-equivalent solution schemes, we demonstrate in Sec. \ref{result} 
a number of pilot numerical applications to analyze the bond dissociation 
profile of strongly correlated systems and justify how our scheme can 
avoid the catastrophic failure of conventional schemes while it
restores significant amount of correlation. Finally we conclude 
our findings in Sec. \ref{conclude}.

\section{Theoretical development}
\label{theory}
\subsection{The structure of scattering operator and the parent iterative n-body Excitation inclusive CCSD (iCCSDn) ansatz:}
\label{dualexp}
As previously mentioned, in the iCCSDn theory, we introduce
an additional \enquote*{scattering} operator that has a quasi-hole/
quasi-particle destruction component and thus it acts on only 
selective excited determinants. 
For example, for a $S_h$ operator labelled by the HF occupied orbitals
$i,j$ and $m$ and virtual orbital $a$ with the associated normal ordered (with 
respect to HF vacuum) operator string $\{a^\dagger m^\dagger ji\}$ acts non-trivially 
on those selected set of determinants in which $i$ and $j$ are occupied while 
$a$ and $m$ are unoccupied. Similar analysis may be done for the $S_p$ operators as well.
Thus in terms of the second quantized operators, the components of 
$S=S_h + S_p$ may be expressed as:
\begin{eqnarray}
S_{h} = \frac{1}{2} \sum_{amij} s^{am}_{ij} \{a^\dagger m^\dagger
ji\} 
\end{eqnarray}
and
\begin{eqnarray}
S_{p} = \frac{1}{2} \sum_{abie} s^{ab}_{ie} \{a^\dagger b^\dagger 
ei\} 
\end{eqnarray}
In general, \textit{a, b, c, ...,e,..} etc. denote the set
of unoccupied particle orbitals and \textit{i, j, k, ...,m,...} etc.
are the set of hole orbitals \textit{with respect to the HF vacuum}. 
Note that in the definition
of $S_{h}$ and $S_{p}$ above, \enquote*{\textit{m}} 
and \enquote*{\textit{e}}
respectively, appear as quasi-hole and quasi-particle \textit{destruction}
operator. Due to the presence of the destruction operators, 
its action on the Hartree-Fock reference determinant is 
trivially zero. Thus $S$ operators satisfy the vacuum 
annihilating condition (VAC): $S_h|\Phi\rangle = 0;\hspace{0.3cm} S_p|\Phi\rangle = 0$. 
The scattering operator acts non-trivially on the excited
functions due to the presence of quasi-hole / quasi-particle
destruction component in it. Thus, the triply excited space
is generated by the action of $S$ operator 
on the doubly excited functions, which is realized 
via the contraction between $S$ and $T_2$ operators.

Following our previous derivation, one may write the correlated 
wavefunction as:
\begin{equation}
|\psi\rangle = \{\exp(S)\}(\exp(T_1 + T_2)|\phi_{HF}\rangle)
\end{equation}
where, $\{...\}$ denotes the normal ordering to avoid the
$S-S$ contraction. Henceforth, we will write $T=T_1+T_2$. One may
further write the above expression as:

\begin{equation}
    |\psi\rangle = \{\exp(S)\}(\sum_{\mu} c_{\mu} |\phi_{\mu}\rangle) 
\end{equation}
where the correlated wavefunction $e^{T}|\phi_{HF}\rangle$ is
expanded in terms of various excited determinants generated by
$T_1$, $T_2$ and their products of various orders. 
Since the contraction among the
$S$ operators are disallowed owing to the normal ordering, all the
$S$ operators are left to act directly on different $\phi_{\mu}$s.
Due to the intrinsic projector in $S$, only 
certain $\phi_{\mu}$s are picked up depending on 
the structure of $S$, for the rest of the $\phi_{\mu}$s
would be annihilated by the action of $S$.
In other words, to have a non-trivial coupling 
between the $S$ and $T$ operators, the destruction 
in $S$ should necessarily get
contracted with the creation operators present 
in $T$. Note that the operators $T$ and $S$ are 
constructed in spinorbital basis, thereby they allow 
all different possibilities of spin orientation for 
both the electrons they involve. In the
present formulation, we would construct the cluster
operators, $T$, in such a way that the resultant 
$\phi_{\mu}$s are adapted to singlet, thus allowing 
$\{\exp(S)\}$ to act on only the singlet-paired 
determinants, leaving out their coupling with the
triplet-paired ones. In the following subsection, we 
would focus on the structure of the cluster operators,
$T$, such that the resulting $\phi_\mu$s are 
singlet-coupled, while the $S$ operators are defined as
above without any further approximation.

\subsection{Choice of the two-body cluster operators:}
\label{singletT}

In this subsection, we elaborate upon the choice of the 
two-body cluster operators. Since the RHF reference function 
includes only the exchange correlation between same spin electrons, it 
treats the same and opposite spin electrons in an 
unbalanced manner. Since the RHF function is biased towards
same spin electrons, correlation energy from any low order
perturbative theory is somewhat biased towards same spin 
electrons. In this section, we approximate the
cluster operator is such a way that the two-electron 
correlation due to triplet combination of the operators,
which include same spin electrons, is entirely
eliminated. Since the two body cluster operator, $T_2$, involves two
electrons, one may build up singlet and triplet paired
components of it by taking suitable linear combinations. 
So, the cluster operator is a combination of these spin
channels: $T_{2} = T_{2}^{sc} + T_{2}^{tc}$ and
these two spin channels can be expressed via the proper 
utilization of symmetric and anti-symmetric pair operators.
Following Scuseria,
\begin{eqnarray}
T_{2}^{sc} = \frac{1}{2} \sum_{ijab} \lambda_{ij}^{ab} X_{ab}^{\dagger} X_{ij}
\hspace{0.6cm}
T_{2}^{tc} = \frac{1}{2} \sum_{ijab} \theta_{ij}^{ab} Y_{ab}^{\dagger} Y_{ij}
\label{eqxx10}
\end{eqnarray}
Here $\lambda$ and $\theta$ are the amplitudes associated 
with $T_{2}^{sc}$ and $T_{2}^{tc}$ respectively. $X$ is the 
symmetric pair operator, and can be 
expressed as:
\begin{equation}
X_{ij} = \frac {1}{\sqrt{2}} (c_{{j_\alpha}} c_{{i_\beta}} + c_{{j_\beta}} c_{{i_\alpha}})
\end{equation}
and $Y$ is the anti-symmetric operator which consists of 
two components. One of the components is for the electrons 
of different spins ($Y^{0}$) and can be denoted as:
\begin{equation}
Y_{ij}^{0} = \frac {1}{\sqrt{2}} (c_{{j_\alpha}} c_{{i_\beta}} - c_{{j_\beta}} 
c_{{i_\alpha}})
\end{equation}
while the other component which takes care of the same 
spin electrons is described as $Y^{\sigma}$:
\begin{equation}
Y_{ij}^{\sigma} = c_{{j_\sigma}} c_{{i_\sigma}}   
\end{equation}
Thus, the complete
expression for the anti-symmetric pair operator $Y$ can be
written as
\begin{equation}
Y_{ab}^{\dagger} Y_{ij} = \sum_{\sigma = \alpha,\beta} (Y_{ab}^{\sigma})^{\dagger} 
(Y_{ij}^{\sigma}) + (Y_{ab}^{0})^{\dagger} (Y_{ij}^{0}) 
\end{equation}
Thus the symmetric pair operator, $X$, takes into account 
the singlet correlation due to of opposite spin electrons,
while the anti-symmetric pair operator $Y$ induces 
correlation due to the anti-symmetric combination of 
opposite spin electrons as well as that of the same spin
electrons. In our approach, we adapted to the singlet
coupling scheme by taking the symmetric combination
of the two-body cluster operators. Thus,
$T_{2}$ is approximated as $T_{2}^{sc}$. One may note that for the same 
spin electrons, the two-body cluster amplitudes follow
the relation: 
${t_{{i_\alpha}{j_\alpha}}^{{a_\alpha}{b_\alpha}}} = -
{t_{{i_\alpha}{j_\alpha}}^{{b_\alpha}{a_\alpha}}}$
and ${t_{{i_\beta}{j_\beta}}^{{a_\beta}{b_\beta}}} = -
{t_{{i_\beta}{j_\beta}}^{{b_\beta}{a_\beta}}}$. Thus the 
symmetric linear combination automatically ensures that 
the contribution from the same spin electrons
gets eliminated, and only the symmetric combination of 
the opposite spin electrons is accounted for.
Starting from a RHF based CCSD theory, one may thus 
redefine the approximated 
cluster operators as $T_{2} = T_{2}^{sc}$, and compute the
corresponding singlet paired cluster amplitudes in each
iteration cycle as:
\begin{equation}
\lambda_{ij}^{ab} = P(ij)\lambda_{ij}^{ab} = P(ab)\lambda_{ij}^{ab}
= P(ij)(ab)\lambda_{ij}^{ab} = \frac {1}{2} (t_{ij}^{ab} + t_{ij}^{ba})
\end{equation}
Here $t_{ij}^{ab}$ denotes the cluster amplitudes 
associated with the excitation from the spatial occupied orbitals 
$i,j$ to the unoccupied orbitals $a,b$, and $P(ij)$ is the 
permutation operator that permutes the orbital labels $i$ and $j$ 
(or $a$ and $b$ for $P(ab)$).
One may note that unlike RPA, all the possible contraction 
channels to determine the cluster amplitudes are taken as 
we have not eliminated any terms from the amplitude equations.
Thus the singlet coupled operators $T_{2}^{sc}$ still follow 
the eigenvector properties as it accounts for the
exact solution of the Schr{\"{o}}dinger equation with the given 
approximation on the wavefunction.

With the approximations discussed, the ansatz comes down to the following:
\begin{equation}
\Omega_{sc} = \{\exp(S)\}\exp(T_{1}+T_{2}^{sc})
\end{equation}
where, $T^{sc} = \frac{1}{2}[t_{ij}^{ab}+t_{ij}^{ba}]
\{{a_{a_\alpha}^{+}a_{b_\beta}^{+}a_{j_\beta}a_{i_\alpha}\}}
=\frac{1}{2} \lambda_{ij}^{ab}\{{a_{a_\alpha}^{+}a_{b_\beta}^{+}a_{j_\beta}a_{i_\alpha}}\}$. 
The Schr{\"o}dinger equation we aim to solve is of the form:
\begin{equation}
H\{e^S\}e^{T_{1}+T_{2}^{sc}}|\phi_{HF}\rangle = E \{e^S\} 
e^{T_{1}+T_{2}^{sc}}|\phi_{HF}\rangle
\end{equation}
So, by replacing $T^{sc}$ in place of $T_2$ we are essentially 
including singlet channel while the two-electron triplet channel is not considered 
on purpose. The opposite spin channels by construction includes
short range interactions while the long range interactions
are missed out due to the elimination of triplet spin channel.
While this might lead to poorer description of bond dissociation, 
the correlation can somewhat be replenished by the inclusion of 
higher rank correlation through the iCCSDn framework. However, due to the restriction 
of the underlying singlet-coupled doubles excitation subspace on which the $S$ acts, 
the resultant triple excited subspace is spanned only by the 
low-spin coupled functions.

While our analysis that follows would assume a closed-shell singlet 
coupled reference function--the HF determinant--in often cases, as shown 
by using geminal functions,\cite{surjan_2015,Li_2003} the mere inclusion of singlet 
coupled functions in the reference may be insufficient for qualitatively correct 
description of multiple bond dissociation. For example, Surja{\'{n}} \textit{et. al.}
demonstrated that one needs to include four-electron singlets by appropriate 
recoupling of triplet coupled geminals for correct description of the 
symmetric stretching of water molecule. In general, even within
a single reference framework, for non-closed-shell cases, some of the triplet 
coupled states are important. However, we have not explored this possibility here.

\subsection{A run-up to the iterative n-body Excitation inclusive CCSD (iCCSDn): Two 
distinct paradigms}

In this subsection, we would first discuss the basic building 
blocks of the parent iCCSDn theory. For this manuscript,
we would proceed with the restriction that the rank-two 
cluster operators are coupled to be singlet-paired; 
however, the mathematical manipulations would remain 
same if the cluster operators are taken to be spin-unrestricted. 
In the context of inclusion of selective low-spin channel
excited determinants towards the correlation, we would design
two distinctly non-equivalent paradigms
to derive the working equations as discussed below.

\subsubsection{Paradigm-I: Explicit projective form with sufficiency condition 
and the genesis of proj-iCCSDn}
\label{projective}
In paradigm-I, we start from the expansion of the correlated wavefunction $|\psi\rangle$:
\begin{equation}
    |\psi\rangle = \{e^S\} e^{ \tilde{T}^{sc}}|\phi_{HF}\rangle
    \label{para1.1}
\end{equation}
where $\tilde{T}^{sc} = T_1 + T_2^{sc}$. Noting the
fact that the wave-operator ansatz includes the product of 
two exponential terms, one may explicitly apply
Wick's theorem between the destruction operators in $S$ and the 
creation operators present in $\tilde{T}^{sc}$
to write Eq. \ref{para1.1} as:
\begin{equation}
    |\psi\rangle = \{e^S e^{\contraction{}{S}{}{\tilde{T}^{sc}} 
    S \tilde{T}^{sc}} e^{\tilde{T}^{sc}}\} |\phi_{HF}\rangle
    \label{para1.2}
\end{equation}
Here the term ${\contraction{}{S}{}{\tilde{T}^{sc}} 
S \tilde{T}^{sc}}$ denotes \textit{all} the possible single, 
double, triple and quadruple contractions between various
powers of $S$ and $\tilde{T^{sc}}$. Note that due to the normal 
ordered structure of $e^S$, there is no explicit 
contraction between two $S$ operators and thus maximum 
of four $S$ operators can contract to the creation 
operators in $\tilde{T}^{sc}$. Here 
${\contraction{}{S}{}{\tilde{T}^{sc}} 
S \tilde{T}^{sc}}$ has the hole-particle structure 
like an excitation operator, the lowest of which starts
from rank-three. Also, note that the terms inside normal
ordering but not explicitly shown to be contracted also 
form an exponential structure, owing to the property of 
an exponential operator. Due to the similar 
hole-particle excitation structure between 
$\tilde{T}^{sc}$ and ${\contraction{}{S}{}{\tilde{T}^{sc}} 
S \tilde{T}^{sc}}$, they commute, and thus Eq. 
\ref{para1.2} can equivalent be written as:
\begin{equation}
     |\psi\rangle = \{e^S e^{\contraction{}{S}{}{\tilde{T}^{sc}}
     S \tilde{T}^{sc} +\tilde{T}^{sc}}\}
     |\phi_{HF}\rangle
    \label{para1.3}
\end{equation}
which can be written in shorthand as
\begin{equation}
     |\psi\rangle = \{e^S\} e^Z|\phi_{HF}\rangle
    \label{para1.32}
\end{equation}
Henceforth, we will no longer carry the complicated 
exponent and we will denote (${\contraction{}{S}{}
{\tilde{T}^{sc}} S \tilde{T}^{sc}}$+ $\tilde{T}^{sc}$ =
Z). Furthermore, the uncontracted $e^S$ and $e^Z$ do not 
have any contraction between them, hence the former acts directly on the $HF$ 
reference leading to annihilate the vacuum. Thus, 
without any loss of generality, Eq. \ref{para1.32} can be
written as:
\begin{equation}
    |\psi\rangle = e^{Z}|\phi_{HF}\rangle
    \label{para1.4}
\end{equation}
Here we have removed the explicit use of the normal 
ordering due to the hole-particle excitation structure
of $Z$. Note that due to the explicit contraction between 
$S$ and $\tilde{T}^{sc}$ in $Z$, the former is
allowed to act only on the singlet-pared determinants
to induce triple and higher excitations. An analysis 
with the spin coupling scheme reveals that such
triply excited determinants are necessarily 
low-spin coupled.
\begin{figure*}[!htbp]
\hglue -0.55cm
        \includegraphics[angle=00,scale=0.87]{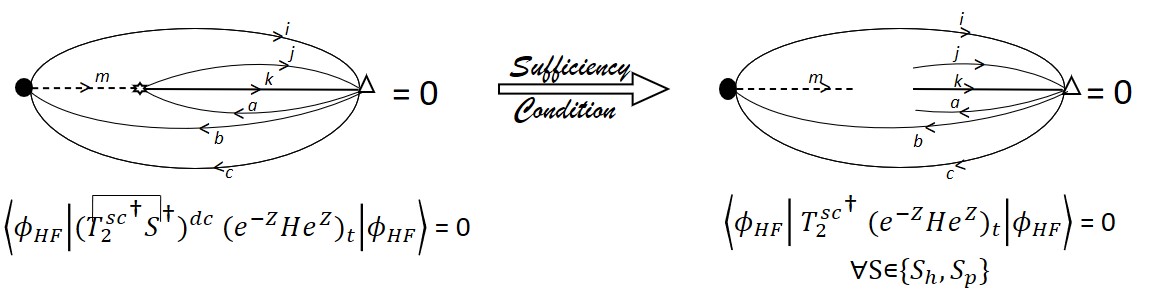}\\
\caption{Diagrammatic representation of the equations that led us to the projective form 
to determine the $s$-amplitudes. Each $s-$amplitude is determined from 
partial (bra-)variation of the sum of three-body residuals (Eq. \ref{para1.6}) 
which are generated via the explicit contractions between several sets of 
rank-three effective hamiltonian elements and the rank-two (singlet-coupled) cluster operators. 
The operator part corresponding to $S$ operators
is not shown explicitly in the right hand figure.}
\label{res100}
\end{figure*}

The Schr\"{o}dinger equation for the correlated ground
state wavefunction now can be written in the usual manner:
\begin{equation}
    He^{Z}|\phi_{HF}\rangle = Ee^{Z}|\phi_{HF}\rangle
\end{equation}
The rank-one and the singlet-coupled rank-two cluster
operators can be determined through explicit projection
against the singly and singlet-coupled doubly excited 
determinants:
\begin{eqnarray}
\langle \chi_s | e^{-Z}He^Z | \phi_{HF}\rangle = 0 \nonumber \\
\langle \chi_d^{sc} | e^{-Z}He^Z | \phi_{HF}\rangle = 0
\label{para1.600}
\end{eqnarray}
Here $\chi_s$ and $\chi_d^{sc}$ are the singly and singlet-coupled 
doubly excited determinants. In the current work, we
have included only the "renormalization terms" (see paradigm-II) 
in the two-body effective hamiltonian, $e^{-Z}He^Z$.
While the rank-one and rank-two cluster operators can be
computed in a straightforward manner, there is no such 
closed form of equation possible for $S$ due to the 
VAC. However, a residual equation 
can be constructed by taking the sum of the bra 
projections of various three-body excitation 
blocks of the effective hamiltonian against rank-three projection 
manifold. This implies that the said residual equation takes the form:
\begin{equation}
\tilde{R}_t = \sum_I\tilde{R}_{t_I} = \sum_I \langle \chi_{t_I} | e^{-Z}He^Z | \phi_{HF}\rangle 
\label{para1.6}
\end{equation}
where $\chi_{t_I}$ is the triply excited determinants with 
collective hole-particle labels as $I$. Since
these triply excited determinants are generated via 
contraction between a number of spin-unrestricted 
$S_J$ and the $T_{2_K}^{sc}$, one may 
expand the bra projection in an effective manner to 
write $\tilde{R}_{t_I}$ that appears in
Eq. \ref{para1.6} as:
\begin{eqnarray}
\tilde{R}_{t_I} = \sum_K \sum_J \langle \phi_{HF}| \contraction[2ex]{} {\left((\contraction{}
    {T_{2_K}^{sc})^\dagger}{}{(S_J)^\dagger}
    T_{2_K}^{sc})^\dagger (S_J)^\dagger\right)_{t_I}}{}{\bigg(e^{-
    Z}He^{Z}\bigg)_{t_I}} \left((\contraction{}{T_{2_K}^{sc})^\dagger}{}{(S_J)^\dagger} 
    T_{2_K}^{sc})^\dagger (S_J)^\dagger\right)_{t_I} \bigg(e^{-Z}He^{Z}\bigg)_{t_I} 
    |\phi_{HF} \rangle
\label{para1.7}
\end{eqnarray}
The indices $K$ and $J$ are restricted to those hyperindices for 
which $\langle \chi_{t_I} | = \langle \phi_{HF}| (\contraction{}{T_{2_K}^{sc})^\dagger}{}{(S_J)^\dagger} T_{2_K}^{sc})^\dagger (S_J)^\dagger$.
Note that although we have constructed this equation 
by explicit projection against the triply excited determinants, 
the unknown parameters are the amplitudes for $T$ and $S$. While the
$t$-amplitudes are solved via Eq. \ref{para1.600}, 
one cannot directly construct a 
projective form of $s$-amplitude determining equation: 
the matrix elements (with respect 
to $\phi_{HF}$) of an effective hamiltonian whose
hole-particle structure is like that of $S$ cannot 
be realized owing to their VAC.
Furthermore, there are more number of equations
corresponding to various three-body projections than 
the number of total $s$-amplitudes which lead to 
redundancy. This implies that there are several set 
of double and triply excited determinants that are 
connected by the same scattering operator. One may 
however, demand that with the prior knowledge of 
$t_{2_K}^{sc}$ (via Eq. \ref{para1.600}), the residual sum 
$\tilde{R}_t = \sum_I \tilde{R}_{t_I}$ be stationary under the first order 
variation of the individual $s-$amplitudes of the 
projection subspace. 
This implies that a partial sum of the
full $\tilde{R}_t$ is set to zero.
This suggests that Eq. \ref{para1.7} holds for each 
$s-$ amplitude, individually. 
This conjecture further implies that 
the quantities satisfying Eq. \ref{para1.7} 
automatically ensure:
\begin{eqnarray}
\sum_I \sum_K \langle \phi_{HF}| \contraction[2ex]{} {\left((\contraction{}
    {T_{2_K}^{sc})^\dagger}{}{(Y_J)^\dagger}
    T_{2_K}^{sc})^\dagger (Y_J)^\dagger\right)_{t_I}}{}{\bigg(e^{-
    Z}He^{Z}\bigg)_{t_I}} \left((\contraction{}{T_{2_K}^{sc})^\dagger}{}{(Y_J)^\dagger} 
    T_{2_K}^{sc})^\dagger (Y_J)^\dagger\right)_{t_I} \bigg(e^{-
    Z}He^{Z}\bigg)_{t_I} |\phi_{HF} \rangle = 0 
\nonumber \\
\hspace{0.5cm}
\forall J \in \left\{S_{h},S_{p}\right\}
\label{para1.8}
\end{eqnarray}
Here $Y_J$ is the operator part of $S_J$ and again the index $K$ is
restricted to certain hyperindices as mentioned above. 
Eq. \ref{para1.8} uniquely determines the corresponding 
$s_J$-amplitudes and the stationary condition 
that leads us from Eq. \ref{para1.7} to Eq. \ref{para1.8} may be considered 
as a physically motivated sufficiency condition. 
This sufficiency condition thus removes the redundancy by ensuring 
each unique $s-$amplitude is determined via a (weighted) sum 
of certain two-body projections on to three-body effective hamiltonian elements 
which are connected by the same operator string of $S$ corresponding to the amplitude. 
A diagrammatic representation of the sufficiency condition is shown in Figure
\ref{res100}. Note that the development of the 
projective scheme above to solve the scattering 
amplitudes is done by restricting all the two-body cluster 
operators (both in the projection manifold and effective hamiltonian) 
to be singlet coupled; however, such a formalism
is entirely general and one does not need to impose any 
additional approximation on the spin-channels to solve 
for generalized excitation operators. In addition to 
the stationarity of $\tilde{R}_{t_I}$ with respect to the individual $s-$amplitudes, 
one may also demand to make the residue of 
Eq. \ref{para1.7} stationary with respect to
the two-body (singlet-paired) cluster amplitudes and as such
this would provide an additional correction term due to the 
nontrivial coupling between $S$ and $\tilde{T}^{sc}$ to 
Eq. \ref{para1.600}. However, this is not considered in the 
current manuscript. The coupling between $S$ and $\tilde{T}^{sc}$
however enters the equation through $Z$.

Note that in the projective form of the working equations, 
one necessarily needs to compute the rank-three 
terms out of $e^{-Z}He^Z|\phi_{HF}\rangle$. The 
explicit expressions of these three-body terms included here are given in the 
supplementary materials.
Since $Z$ contains one, (singlet coupled) two- and implicit three-body
excitation operators, the effective Hamiltonian $e^{-Z}He^Z$ can be done in a similar 
manner one performs CCSDT. However, one just has to take the singlet combination of $T_2$ 
at each step and interpret each element of $T_3$ as: $T_{ijk}^{abc} \leftarrow  
(\sum\limits_{m} S_{ij}^{am} (T^{sc})_{mk}^{bc} + \sum\limits_{e} S_{ie}^{ab} 
(T^{sc})_{jk}^{ec})$. Clearly, this involves $N^8$ computational 
scaling. However, the number of amplitudes that one determines in this 
formulation is still dictated by the number of $T_1$, 
(singlet-coupled) $T_2$ and $S$ operators, which 
together are order of magnitude less than the 
usual CC formulation with single, double and triple 
excitations. One may further note that one can 
induce quadruple and even higher excitations with 
increased number of matrix operations, however, the 
number of variables in such cases also do not exceed any 
further. We would generically denote the projective version of iCCSDn as proj-iCCSDn.
The particular variant of proj-iCCSDn where only low spin channels
are incorporated will be denoted as proj-iCCSDn-LS.

\subsubsection{Paradigm-II: amplitude formulation with many-body expansion and the 
genesis of iCCSDn-LS}
\label{amplitude}

While the projective formulation discussed above is 
theoretically appealing, it involves the explicit 
construction of rank-three diagrams. In the following
paradigm-II, we will deal with explicit operator 
formulation where we would construct the residue 
for various hole-particle sectors to demand their 
amplitudes vanish at the solution. This would involve 
many-body expansion, as advocated by Nooijen\cite{nooijen2000}, and
we would refer this to as the \textit{amplitude formulation}.

\begin{figure*}[!htbp]
\hglue -0.75cm
        \includegraphics[angle=00,scale=0.1]{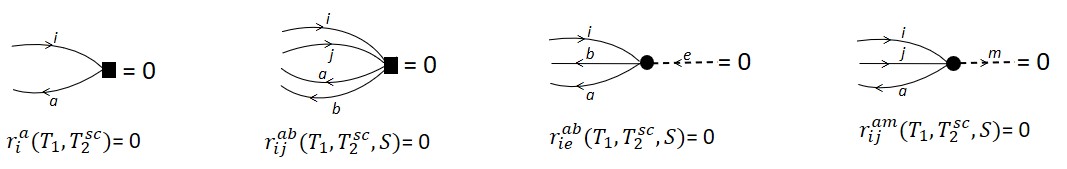}\\
\caption{Diagrammatic representation of the amplitude equations to determine 
the cluster and scattering amplitudes. The singlet-coupled cluster amplitudes 
are determined by taking their symmetric combination at each step and the various 
residues are constructed with the singlet-coupled cluster amplitudes.}
\label{res101}
\end{figure*}

\begin{figure*}[!htbp]
\hglue -0.9cm
        \includegraphics[angle=00,scale=0.85]{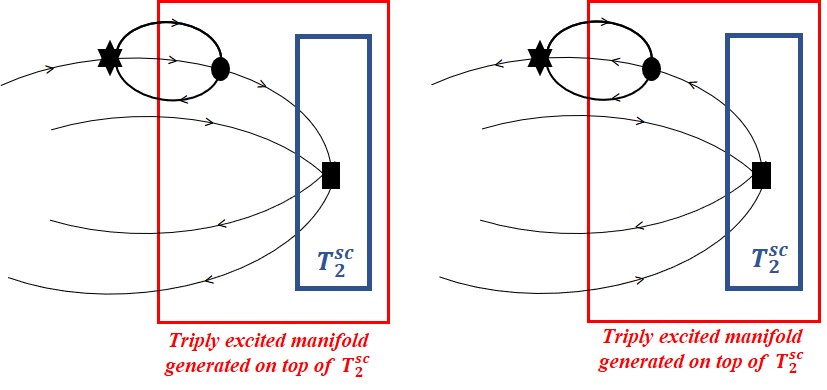}\\
\caption{Representative diagrams of the renormalization term where 
$S$ operator (denoted by the filled circle) gets coupled with $T_2^{SC}$ 
operators (denoted by filled squares) and eventually contributes to the 
equation of $T_2^{SC}$. The black rectangles denote the intermediate 
low spin coupled triply excited states. The filled stars denote the 
hamiltonian matrix elements.}
\label{res102}
\end{figure*}

Towards the amplitude formulation of the iCCSDn working
equations, we proceed taking a different route staring 
from our ansatz as presented in Eq. \ref{para1.1}.
However, unlike the projective formulation, one does
not have to explicitly construct the three-body 
excitation blocks; rather, one may work entirely in 
the amplitude space to construct an effective hamiltonian.
Noting the fact that the similarity transformation 
of a normal ordered exponential operator is 
non-trivial, one may define an effective 
(similarity transformed) hamiltonian, $W$ such that 
\begin{equation}
\{e^S\} W = H \{e^S\}
\label{eqxx4}
\end{equation}
holds. With the help of Wick's theorem and exploiting 
the properties of an exponential operator, one may 
design a recursive equation for $W$ as follows:
\begin{eqnarray}
W=\{\contraction{}{H}{}{e^S} H e^S\}  - \{\contraction[2ex]
{}{(e^S-1)}{}{\contraction{}{H}{}{e^S} H e^S} (e^S-1)
\contraction{}{H}{}{e^S} H e^S\} +
\{\contraction[3ex]{}{(e^S-1)}{}{\contraction[2ex]
{}{(e^S-1)}{}{\contraction{}{H}{}{e^S} H e^S} (e^S-1)
\contraction{}{H}{}{e^S} H e^S} (e^S-1) \contraction[2ex]
{}{(e^S-1)}{}{\contraction{}{H}{}{e^S} H e^S} (e^S-1)
\contraction{}{H}{}{e^S} H e^S \}  - \cdot\cdot\cdot
\label{eqxx6}
\end{eqnarray}
We refer the readers to our earlier 
publications\cite{iccsdn1,maitra_correlation_2017,maitra_coupled_2017}
for the details of the derivation. In our applications,
we have truncated $W$ after the second term on the 
right hand side of Eq. \ref{eqxx6}. Furthermore, $W$ 
is a many-body operator in general, and one may thus 
develop various approximate schemes with different 
levels of sophistication by retaining up to certain 
rank of terms in $W$. We emphasize that in all
the terms generated in $W$ (irrespective of its 
many-body rank), there is \textit{at least}
one destruction operator arising out of the contraction
between the hamiltonian and $S$ from the right. 
With the similarity transformed hamiltonian
$W$, one may cast the Schr{\"o}dinger equation as:
\begin{equation}
 W e^{\tilde{T}^{sc}}|\phi_{HF}\rangle = E 
 e^{\tilde{T}^{sc}}|\phi_{HF}\rangle   
\end{equation}
which is akin to the conventional CC equations with 
the hamiltonian replaced by the similarity 
transformed hamiltonian. Expanding $e^{-\tilde{T}^{sc}}We^{\tilde{T}^{sc}}$ 
via the Baker-Campbell-Hausdorff expansion in terms 
of the explicitly connected quantities, one may note
that only those terms survive in the hole-particle 
excitation structures which have explicit
contraction between $S$ and $\tilde{T}^{sc}$. This
in turn takes care of the connected triple excitations 
where the $S$ operators act selectively on the 
singlet-coupled doubly excited determinants. In the
following, we would discuss about the approximations
made in the similarity transformed hamiltonian $W$
and also would discuss about the many-body 
expansion of the quantity $\{\contraction{}{W}{}{e^{\tilde{T}^{sc}}} W 
e^{\tilde{T}^{sc}}\}$ that leads to 
various amplitude equations.

To reduce the computational overhead and at the same
time to incorporate the leading order contribution to
connected rank-three excitations, we would keep 
minimal number of terms in $W$ from the expansion
Eq. \ref{eqxx6}. Note that in $W$, the leading order 
one-body term is the unperturbed Fock operator and 
the leading two-body term is the two-body interaction 
$V$. However, the expansion also generates three-body 
and higher rank terms, which we neglect entirely. 
Thus the similarity transformed hamiltonian is 
restricted to have only one and two-body terms. 
Furthermore, the two-body terms in $W$ have been 
approximated at its leading order where only $V$ is
incorporated. For the hole-hole and
particle-particle sectors of the one-body terms 
in $W$, we incorporate additional terms arising out 
of the contraction between $V$ and $\{e^S\}$:
$\{\contraction{}{V}{}{e^S} V e^S\}$, the higher rank
terms of which are entirely neglected. This 
approximation automatically restricts the expansion
of Eq. \ref{eqxx6} to the first term on the right hand 
side. While the two-body component of $W$ is 
approximated through the two-body integrals, the leading order of
the contribution of $S$ is folded into the one-body \textit{dressed} 
Fock matrix element through a set of \textit{renormalization terms} 
(vide infra). We refer our earlier publication for the details\cite{iccsdn1}.

Note that in our scheme, one does not explicitly 
determine the triple excitation amplitudes;
rather one determines $s$ amplitudes in a coupled
manner along with the $t_1$ and $t_2^{sc}$ amplitudes
from their respective residue equations.
The strength of Fock space 
formulation is that we can think in terms of operators 
without considering the determinants. Thus, without bothering about 
the wavefunction, one may employ the many-body expansion of the 
double similarity transformed hamiltonian
$R=e^{-\tilde{T}^{sc}}We^{\tilde{T}^{sc}}$, to  
write:
\begin{equation}
R  = r_0 + r_p^q\{E_q^p\} + \frac{1}{4} r_{pq}^{st}\{E^{pq}_{st}\} + \frac{1}{36}
r_{pqs}^{tuv}\{E^{pqs}_{tuv}\} + ...
\label{eqxx17}
\end{equation}
where $p,q,r,s,...$ are general (hole or particle)
orbital indices. 
Clearly, the above many-body 
expansion contains all $N$-body terms with all 
possible hole-particle scattering structures. 
Following Nooijen.\cite{nooijen2000}, the 
corresponding $t-$ and $s-$ amplitudes are 
obtained by demanding the associated amplitudes
vanish:
\begin{equation}
r^{a}_{i} = r^{ab}_{ij} = r^{am}_{ij} = r^{ab}_{ie} = 0
\label{eqxx18}
\end{equation}

While the first two terms are used to determine 
the one- and the two-body cluster amplitudes, the 
last two terms determine $s_h$ and $s_p$ amplitudes,
respectively. A diagrammatic representation of the 
amplitude equations, Eq. \ref{eqxx18} is shown in 
Figure \ref{res101}. Also, a typical \textit{renormalization}
term that includes the nontrivial coupling between 
$T_2^{sc}$ and $S$ is shown in Figure \ref{res102}. 
Although we had taken the symmetric pair
operator to include only the singlet-paired component 
of $T_2$ operators, the corresponding two-body residue 
$R_{ij}^{ab}$ may contain all possible spin orientations.
Following the determination of the cluster amplitudes
from these residues, one still needs to take the symmetric
combination to determine the corresponding singlet
paired amplitude $\lambda$. This is repeated at each step.

Note that in the amplitude determining
equations, both $T$ and $S$ operators are treated on 
the same footings and as such, both are iteratively 
determined from their coupled nonlinear equations. We
mention though that in the residue equations for the
$s_h$ and $s_p$ amplitudes, no nonlinear terms were 
included. Also, the $s-$amplitudes are determined 
via a set of \textit{local} denominators. 
The leading order of these $s-$amplitudes 
can be computed as: $[H_0 , S^{(1)}] + V = 0$, 
resulting in the following perturbative structure:
\begin{equation}
s_{ij}^{am}=\frac{v_{ij}^{am}}{(f_{ii}+f_{jj}-f_{aa}-f_{mm})}
\label{eqyy19}
\end{equation}
and
\begin{equation}
s^{ab}_{ie} = \frac{v^{ab}_{ie}}{(f_{ii}+f_{ee}-f_{aa} - f_{bb})}
\label{eqxx19}
\end{equation} 
In the amplitude formulation of iCCSDn, the 
determination of the $s-$amplitudes adds a computational
overhead of $n_{v, act}n_o n_v^4$ on top of the usual 
scaling of $n_o^2 n_v^4$ that is required to determine 
the amplitudes corresponding to the $T_2$ operators. 
Thus, the overall scaling of iCCSDn is marginally 
more than the conventional CCSD, but 
never exceeds $N^6$.
We would denote the amplitude formulation 
of iCCSDn with low-spin coupled excitations
as iCCSDn-LS.

\section{Result and Discussion}
\label{result}
In this section, we demonstrate the efficacy of our 
method in handling molecular strong correlation. 
We have taken some strongly correlated 
systems which are known to be challenging for
any single reference based electronic structure 
method. In these cases, the Potential Energy
Surfaces obtained from our scheme 
are compared to other known methods including the 
singlet paired CCSD (CCSD0). It is shown that 
our method performs quite well throughout the
weakly and strongly correlated regions of
molecular geometry without any catastrophic failure
yet recovering substantial amount of dynamic correlation.
All the calculations were done using in-house software
platform which is interfaced to PySCF\cite{pyscf} that 
generates the integrals and orbitals.

\subsection{Potential energy curve for $N_2$ dissociation:}
\begin{figure}[!htbp]
\hglue -1.10cm
        \includegraphics[angle=00,scale=0.65]{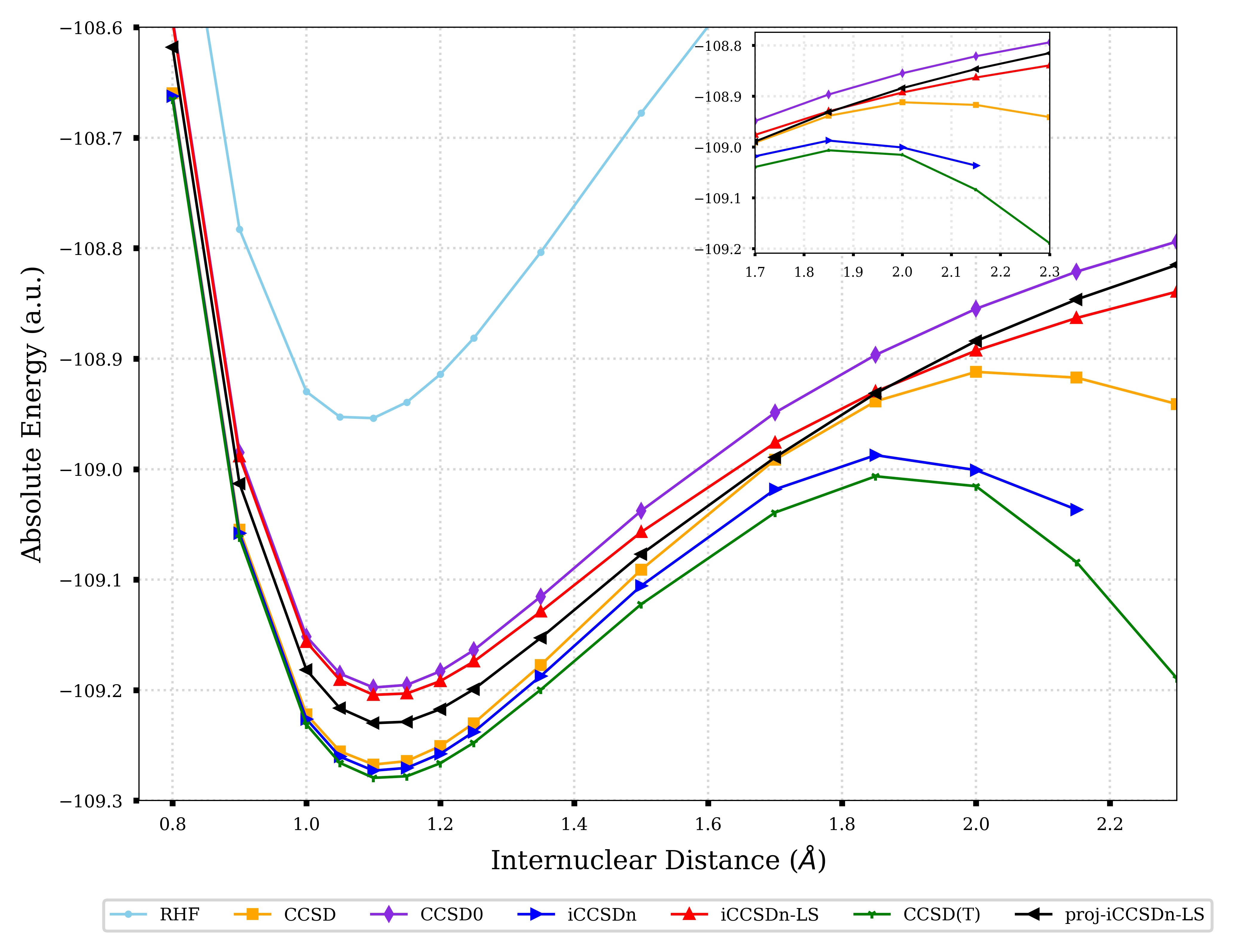}
\caption{Potential Energy Surface (PES) of $N_2$ molecule in cc-pVDZ basis.
The low-spin paired CC methods (e.g. CCSD0, iCCSDn-LS, 
proj-iCCSDn-LS) produce the entire potential surface with qualitative accuracy while 
the conventional methods show catastrophic failure.}
\label{res7}
\end{figure}

The dissociation profile of $N_2$
molecule is known to be one of the most difficult 
test cases to check the efficacy of any newly developed electronic 
structure method. The molecule, even at its equilibrium 
geometry, shows the signature of strong electronic 
correlation. Figure \ref{res7} shows the dissociation
curve of $N_2$ with respect to intermolecular distance 
in cc-pVDZ basis. In the dissociation limit, CCSD 
and CCSD(T) shows catastrophic break down, as the
potential energy curves turn over. The same applies
for the parent iCCSDn method as well. With the two 
body cluster amplitudes restricted solely to the 
singlet coupled channel, the CCSD0 method seems to 
circumvent this problem qualitatively, although it 
is not quantitatively very accurate even at the
equilibrium geometry. The amplitude formulation of
iCCSDn with the two-body excitation manifold restricted
to the singlet channel (the iCCSDn-LS) recovers the 
lost correlation of CCSD0 in the dissociation limit
by stabilizing the energy without any catastrophic
failure. However, it captures little correlation in 
the regions of molecular equilibrium geometry. 
The proj-iCCSDn-LS, on the other hand, 
stabilises the energy over CCSD0 uniformly throughout 
the entire potential energy profile. However, 
irrespective of the solution strategy, one
must include the triplet channel contribution to
energy for the quantitative accuracy.

\subsection{Potential energy curve for linear $H_6$ dissociation:}
\begin{figure}[!htbp]
\hglue -0.75 cm
         \includegraphics[angle=00,scale=0.65]{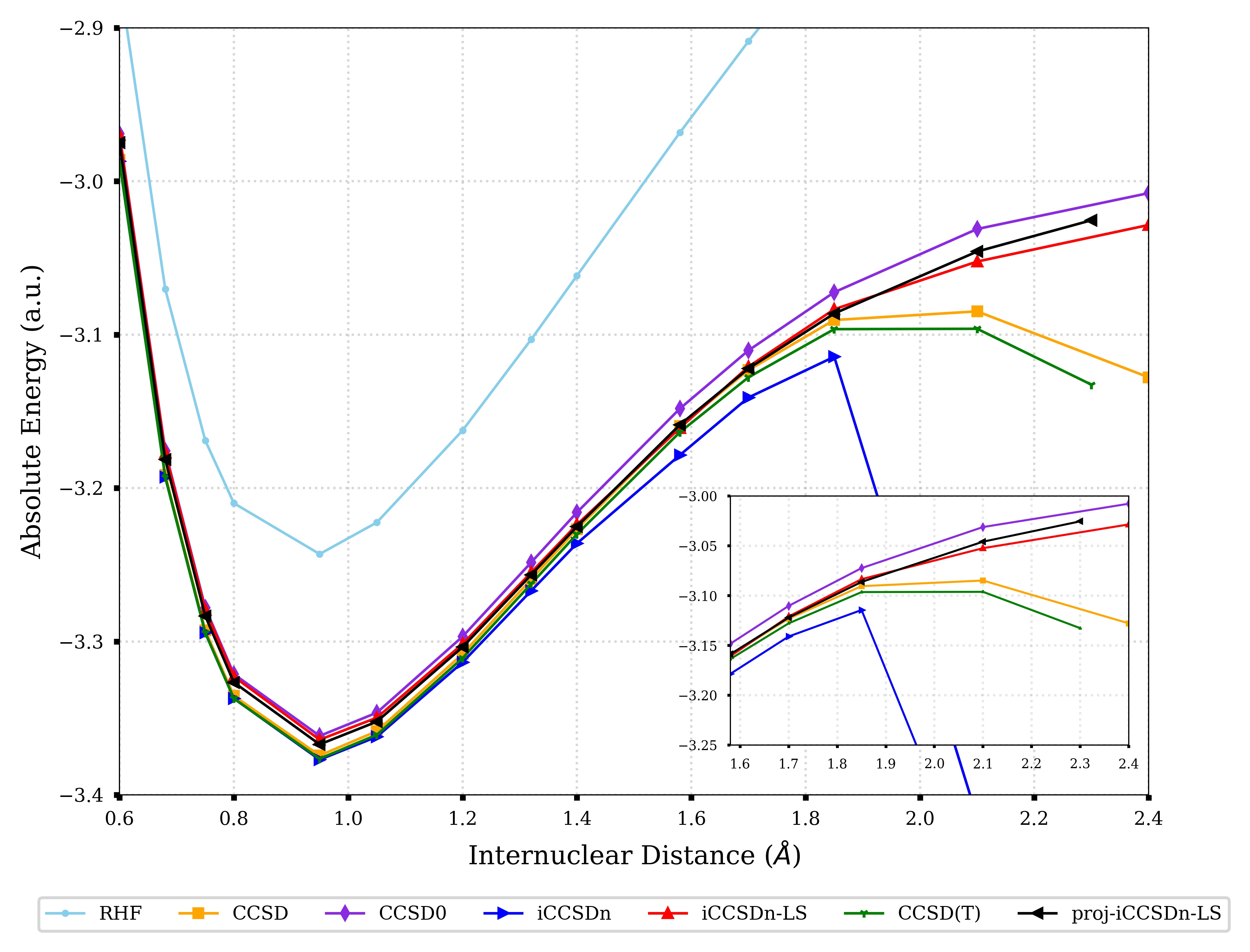}
\caption{PES of linear $H_6$ molecule in cc-pVDZ basis. Unlike 
the conventional methods, the methods with low spin channel coupling 
i.e. CCSD0, iCCSDn-LS and proj-iCCSDn-LS recover the qualitatively correct 
dissociation behavior. }
\label{res8}
\end{figure}

Simultaneous dissociation of six equally spaced linear H 
atoms in $H_6$ is another interesting problem where
the conventional CC methods fail poorly. Figure \ref{res8} 
depicts the PES for linear $H_6$ molecule dissociation
with respect to the interatomic distances in the 
cc-pVDZ basis. Similar to that of previous case of $N_2$,
the conventional CCSD with full singles and doubles
excitation space, as well as the parent iCCSDn fail to
predict the molecular dissociation: after about 
1.8 {\AA}, the iCCSDn methods breaks down steeply, 
while the same happens for CCSD after 2.0 {\AA}. CCSD(T) 
is also entirely plagued in the dissociation limit as 
it shows the most severe catastrophic turn over. 
The singlet paired schemes, namely the CCSD0, iCCSDn-LS
and proj-iCCSDn-LS recover the 
correct qualitative behavior. 
Around the equilibrium geometry, the proj-iCCSDn-LS
is slightly better than the amplitude counterpart while
the opposite behavior is observed in the highly 
stretched geometry. Both the iCCSDn-LS and proj-iCCSDn-LS
variants perform somewhat better than CCSD0 
owing to the fact that the former is able to recover 
part of dynamical correlation via implicit triple
excitations. The important point here is to note that 
the restriction on the spin-channel at the CCSD level or 
the iCCSDn methodology built on top, can qualitatively 
explain the dissociation of strongly correlation systems 
like linear $H_6$ model.

\subsection{Symmetric stretching of $H_2O$}
Symmetric stretching of water molecule is another
difficult test case for electronic structure methodology
as it warrants multi-reference description over the 
potential energy surface. We have systematically 
studied its potential energy profile with iCCSDn-LS and proj-iCCSDn-LS, and 
compared and contrasted our results to the spin-channel
unrestricted counterparts. Due to the availability of the reference 
FCI results, we have chosen to work with the STO-3G 
basis.

\begin{figure}[!htbp]
\hglue -0.8 cm
         \includegraphics[angle=00,scale=0.65]{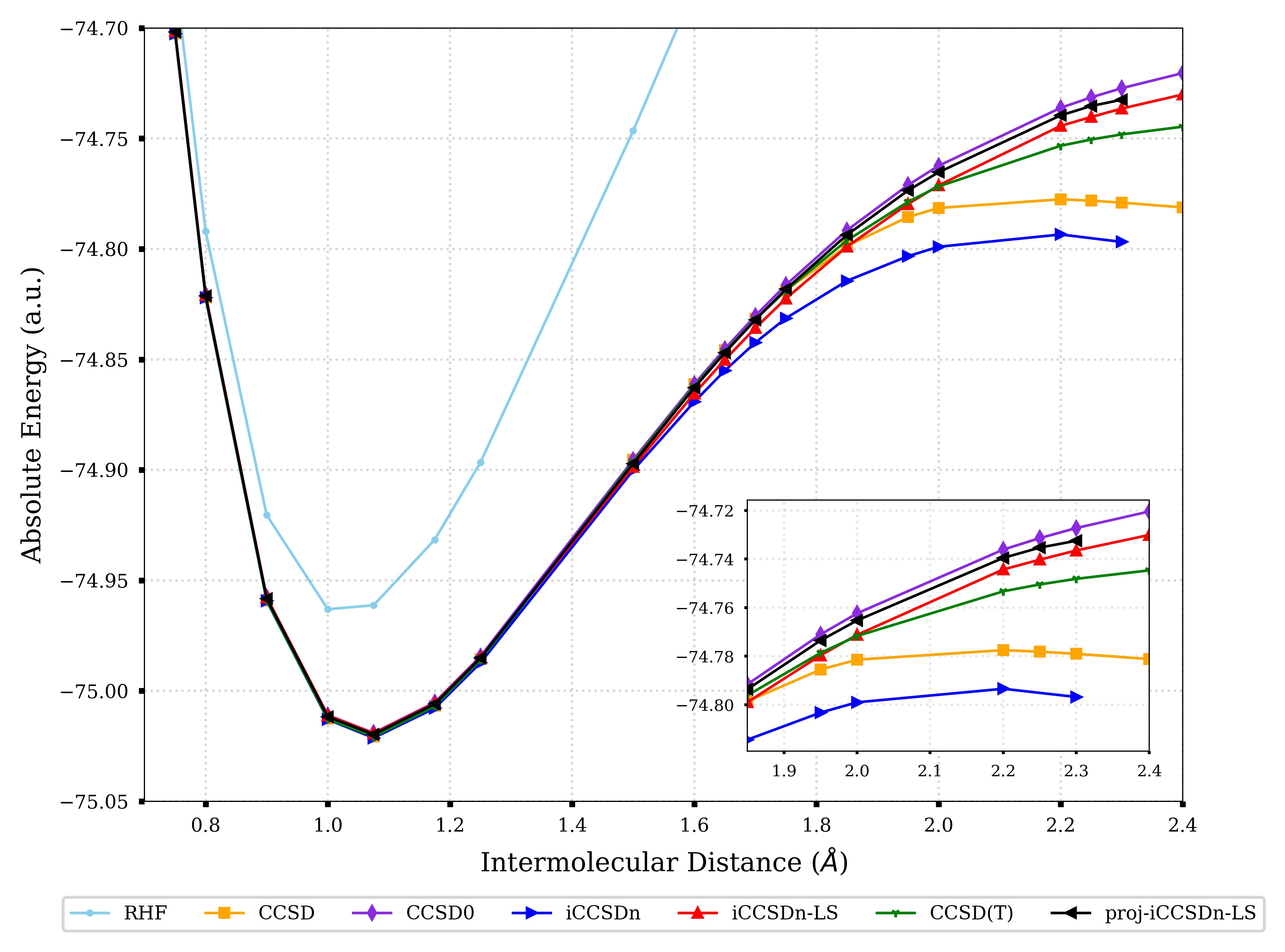}
\caption{PES for symmetric stretching of water molecule using STO-3G basis set. 
Similar conclusion may be made as the cases discussed previously.}
\label{res12}
\end{figure}

As in the previous examples, the CCSD and iCCSDn tend 
to turn over when the $O-H$ bonds are sufficiently 
stretched (beyond $R_{O_H}=2.0$ \AA) as shown in 
Figure \ref{res12}. All the methods that exploit the 
singlet channel description of the two-body cluster
operators fix this problem. Both the iCCSDn-LS as well as
proj-iCCSDn-LS variants show improvement over CCSD0 in the 
dissociation limit, although they are hard to distinguish
around the equilibrium geometry in the scale presented.
This is due to the weak correlation effects around the
equilibrium region that the importance of triples is 
not entirely dominant. However, the well-behaved 
nature of the potential energy surface in the 
dissociation regime for all the singlet-paired schemes
reinforces the importance of spin-channel restriction
for qualitative accuracy.

\subsection{Circular $H_4$}
\begin{figure}[!htbp]
\hglue -1.3 cm
         \includegraphics[angle=00,scale=0.65]{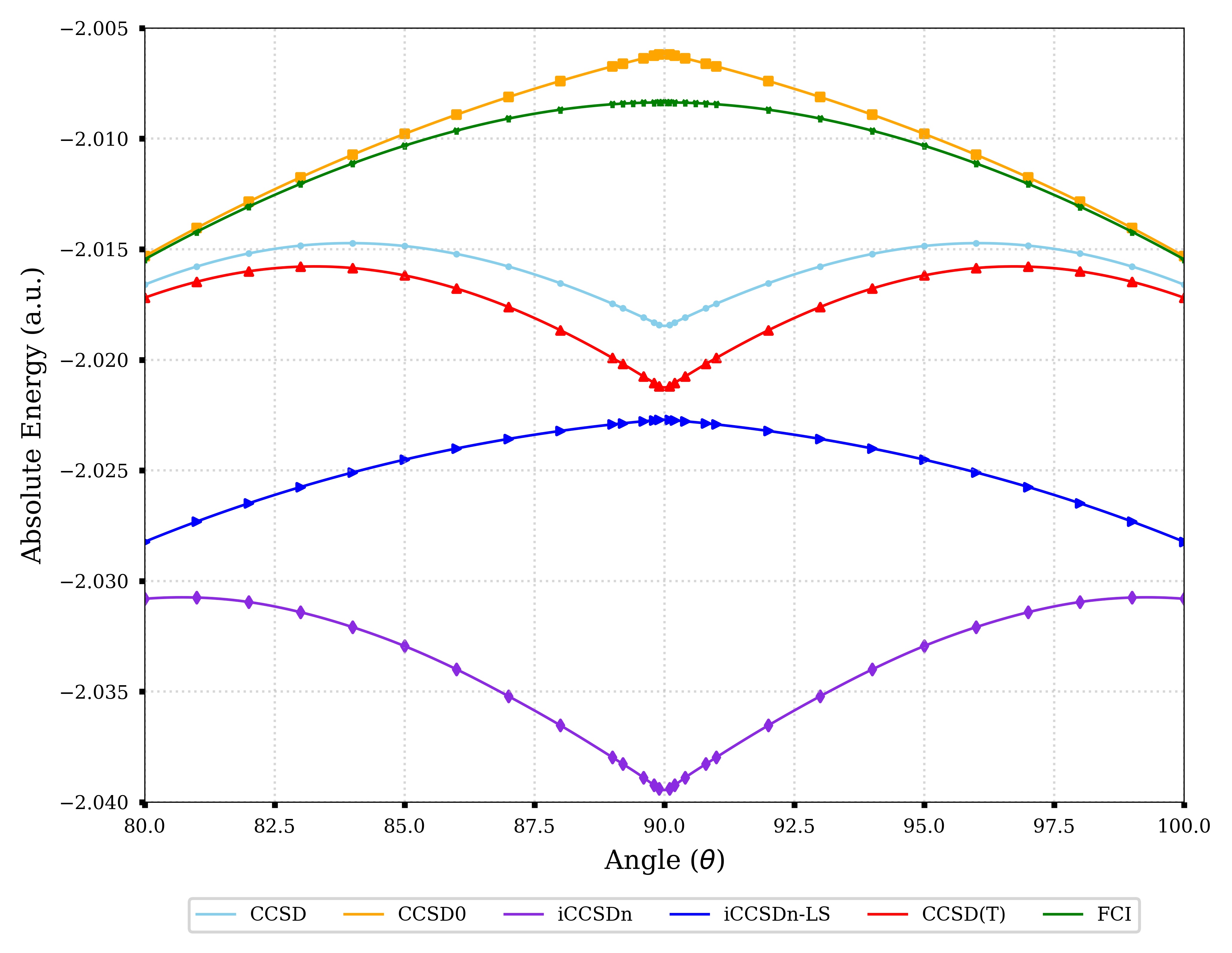}
\caption{PES of Circular $H_4$ molecule in cc-pVDZ basis.
iCCSDn-LS scheme, though not quantitatively
accurate, shows correct qualitative behavior throughout the PES. Furthermore, it 
fixes non-parallelity and kink at $\theta$ = 90$\textdegree$.}
\label{res10}
\end{figure}

$H_4$ on a circle is known to cause problems for standard CC. 
The model consists of four Hydrogen atoms on a circle
of radius R = 1.738 Å, separated by an angle $\theta$ as shown 
in Figure. \ref{res10}. For small and large values of $\theta$, 
the system can approximately be described by two non-interacting
$H_2$ molecules and can well be described by single reference 
methods like CCSD. As $\theta$ approaches 90\textdegree, the system 
acquires a two-fold degeneracy which results in strong correlation.

Figure \ref{res10} shows the results for CCSD, iCCSDn and 
allied methods in cc-pVDZ basis for $\theta$ between
85\textdegree and 95\textdegree. Clearly these methods shown distinct cusps around
90\textdegree. In fact, the CCSD plot is qualitatively 
different to that of FCI. Inclusion of connected triple excitations,
either perturbatively via CCSD(T) or implicitly via iCCSDn, fails 
to rectify this. CCSD0 on the other hand, corrects the qualitative 
inaccuracy of CCSD, although shows a definite cusp
at the crossover point. Although iCCSDn-LS overshoots correlation 
energy throughout the potential energy surface, it shows clear improvement 
over CCSD0 at the strong correlation region which is evident 
from the smooth maxima as well as
the non-parallelity with respect to the FCI. 

While Proj-iCCSDn-LS is demonstrated to be more accurate that the 
iCCSDn-LS counterpart, it was observed that for highly stretched geometries of 
$H_6$ and $H_2O$ as well as for the strongly correlated regime for $H_4$, the former
becomes numerically unstable. As such for these cases, the residual norm for 
Proj-iCCSDn-LS could not be converged to the same level of accuracy as other points 
and hence these results are not indicated beyond a certain bond lengths.

\section{Summary and future direction}
\label{conclude}
In this manuscript, we have explored the methodology of
inclusion of the implicit triple excitation on top of the
singlet paired CCSD theory. While the singlet paired 
CCSD is well known in the literature for quite sometime,
this is the first time we explore the qualitative and 
quantitative accuracy of the dissociation profile of 
various strongly correlated molecules with the implicit
inclusion of high rank excitations. With a sequential 
exponential ansatz, since the scattering operators act 
on the doubly excited determinants to span the triply
excited manifold, 
the restriction on the spin-channel at the doubles
automatically ensures the triply excited space spanned  
only by the corresponding low-spin determinants. 
Towards this along with the hitherto developed amplitude 
formulation, we, for the first time, introduced a novel 
solution strategy through projective equations for the
amplitudes involving generalized hole and particle 
orbitals, bypassing the redundancy through physically 
meaningful sufficiency conditions. In the projective 
formulation, the amplitudes of the scattering operators
are determined by projecting against the triply 
excited determinants which is lowest in excitation 
rank that the scattering operators lead to by their action
on the doubly excited determinants and followed by 
imposing suitable sufficiency conditions. 
In our singlet coupled scheme, we restricted such
triple projection manifold to be spanned by certain
low-spin determinants. However, the projective
formalism thus developed is completely general for any 
arbitrary two-body operator with effective excitation
rank of one. An application of this for strongly 
correlated molecules is currently underway. 
While the amplitude and the projective formulations
are characteristically different, they have the exactly same number 
of unknown parameters. However, a 
comparative analysis of the effective hamiltonian elements 
between these two approaches may reveal interesting
feature about the near optimal parameter space. This is currently 
being carried out in our group. As such, 
both the amplitudes and projective formulations, with 
the appropriate restrictions on the spin channels, 
are shown to be qualitatively accurate and the resulting
methodologies recover substantial high-rank dynamical 
correlation towards quantitative accuracy. 
It would be interesting to explore the importance of triplet
spin coupled excitations for general non-closed-shell references. 
This will be explored in near future.

\section*{Supplementary Material}

See the supplementary material for the algebraic expressions for the
three-body terms that appear in Eq.\ref{para1.6}.

\section*{Acknowledgements}
The authors thank Professor Debashis Mukherjee, India for 
many stimulating discussions during the development of 
the theory. AC thanks Industrial Research and Consultancy 
Center (IRCC), IIT Bombay, for research fellowship. 
Constructive criticisms from an anonymous reviewer is 
gratefully acknowledged.

\section*{Author Declarations}
\subsection*{Conflict of Interest:}
The authors have no conflict of interest to disclose.

\subsection*{DATA AVAILABILITY}
The data that support the findings of this study are
available from the corresponding author upon 
reasonable request.

\end{document}